\documentclass[a4paper]{aa}
\usepackage{txfonts}

\usepackage{url}
\usepackage{graphicx}
\usepackage[hyphenbreaks]{breakurl}
\usepackage{float}
\usepackage{subfig}
\usepackage{multirow}
\usepackage{amsmath}
\usepackage{hyperref}

\outer\def\gtae {$\buildrel {\lower3pt\hbox{$>$}} \over 
{\lower2pt\hbox{$\sim$}} $}
\outer\def\ltae {$\buildrel {\lower3pt\hbox{$<$}} \over 
{\lower2pt\hbox{$\sim$}} $}

\begin{document}

\title{Searching for outbursts from Symbiotic Binaries in GOTO and ATLAS data}

\subtitle{}

\author{G. Ramsay\inst{1}
\and K. Ackley \inst{2}
\and S. Belkin \inst{3}
\and P. Chote \inst{2}
\and D.  Coppejans \inst{2}
\and M. J. Dyer \inst{4,5}
\and R. Eyles-Ferris \inst{5}
\and B. Godson \inst{2}
\and D. Jarvis \inst{4}
\and Y. Julakanti \inst{2}
\and L. Kelsey \inst{6}
\and M. R. Kennedy \inst{7}
\and T.~L. Killestein \inst{2}
\and A. Kumar \inst{8,2}
\and A. Levan \inst{9,2}
\and S. Littlefair \inst{4}
\and J. Lyman \inst{2}
\and M. Magee \inst{2}
\and S. Mandhai \inst{10}
\and D. Mata S\'anchez \inst{11}
\and S. Mattila \inst{12,13}
\and J. McCormac \inst{2}
\and J. Mullaney \inst{5}
\and D. O'Neill \inst{14,2}
\and C. Phillips \inst{2}
\and K. Pu \inst{3}
\and M.  Pursiainen \inst{2}
\and A. Sahu \inst{2}
\and S. Moran \inst{15}
\and M.  Shrestha \inst{3}
\and E. Stanway \inst{2}
\and R. Starling \inst{15}
\and Y. Sun \inst{15}
\and K. Ulaczyk \inst{2}
\and L. Vincetti \inst{1,16,17}
\and B. Warwick \inst{2}
\and E. Wickens \inst{18}
\and K. Wiersema \inst{19}
\and D. Steeghs \inst{2}
\and D. K. Galloway \inst{3,20}
\and V. S. Dhillon \inst{4,11}
\and P. O'Brien \inst{15}
\and K. Noysena \inst{21}
\and R. Kotak \inst{12}
\and R. P. Breton \inst{14}
\and L. K. Nuttall \inst{18}
\and B. Gompertz \inst{14}
\and J. Casares\inst{11}
\and D. Pollacco \inst{2}
}

\institute{Armagh Observatory \& Planetarium, College Hill, Armagh, BT61 9DG, UK\label{inst1}\and 
Department of Physics, University of Warwick, Coventry, CV4 7AL, UK\label{inst2}\and
School of Physics \& Astronomy, Monash University, Clayton VIC 3800, Australia\label{inst3}\and
Astrophysics Research Cluster, School of Mathematical and Physical Sciences, University of Sheffield, Sheffield S3 7RH, UK\label{inst4}\and
Research Software Engineering, University of Sheffield, Sheffield, S1 4DP, UK\label{inst5}\and
Institute of Astronomy and Kavli Institute for Cosmology, University of Cambridge, Madingley Road, Cambridge CB3 0HA, UK\label{inst6}\and
School of Physics, University College Cork, Cork, T12 K8AF, Ireland\label{inst7}\and
Centre for Electronic Imaging, The Open University, Walton Hall, Milton Keynes MK7 6AA, UK\label{inst8}\and
Radboud University, Postbus 9010, 6500 GL, Nijmegen, Netherlands\label{inst9}\and
Jodrell Bank Centre for Astrophysics, Department of Physics and Astronomy, University of Manchester, Manchester M13 9PL, UK\label{inst10}\and
Instituto de Astrof\'isica de Canarias, E-38205 La Laguna, Tenerife, Spain\label{inst11}\and
Department of Physics \& Astronomy, University of Turku, Vesilinnantie 5, Turku, FI-20014, Finland\label{inst12}\and
School of Sciences, European University Cyprus, Diogenes Street, Engomi, 1516 Nicosia, Cyprus\label{inst13}\and
School of Physics and Astronomy, University of Birmingham, Birmingham B15 2TT, UK\label{inst14}\and
School of Physics \& Astronomy, University of Leicester, University Road, Leicester, LE1 7RH, UK\label{inst15}\and
School of Physics, Trinity College Dublin, College Green, Dublin 2, Ireland\label{inst16}\and
Astronomy \& Astrophysics Section, DIAS Dunsink Observatory, Dublin Institute for Advanced Studies, Dublin, D15 XR2R, Ireland\label{inst17}\and
Institute of Cosmology and Gravitation, University of Portsmouth, Portsmouth, PO1 3FX, UK\label{inst18}\and
Centre for Astrophysics Research, University of Hertfordshire, Hatfield, AL10 9AB, UK\label{inst19}\and
Institute for Globally Distributed Open Research and Education (IGDORE)\label{inst20}\and
National Astronomical Research Institute of Thailand, Chiang Mai 50180, Thailand\label{inst21}\and
Departamento de Astrof\'isica, Univ. de La Laguna, E-38206 La Laguna, Tenerife, Spain\label{int22}\\
\email{gavin.ramsay@armagh.ac.uk}
}

\date{Accepted 6 Oct 2026}

\abstract {Symbiotic Binaries contain a white dwarf accreting material
  from a red giant star typically through a wind. We present the
  results of a search for outbursts from Symbiotic Binaries using
  photometric data obtained using the GOTO all-sky survey taken from
  2023 onwards. After identifying ten candidate outbursting systems,
  we used ATLAS photometry to characterise their photometric behaviour
  before 2023, leaving five systems which showed photometric behaviour
  consistent with an outburst. The ATLAS data also provides colour
  information which gives insight to the nature of the outburst. The
  outburst from LMC N67 is the first reported Z And-type outburst from
  a Symbiotic binary in the LMC. In OGLE SMC-LPV-4044 and HK Sco we
  find previously unreported outbursts. QW Sge and V4141 Sgr show
  outbursts starting in 2024, which have already been reported and are
  ongoing. By better identifying and characterising Z And-type
  outbursts from many systems, it will be possible to better
  understand the physics of these events, which are still not fully
  understood.}

\keywords{Physical data and processes: Accretion, accretion disks:
  binaries: symbiotic – novae: stars: individual: LMC N67, OGLE
  SMC-LPV-4044, HK Sco, QW Sge, V4141 Sgr.}

\maketitle

\section{Introduction}

Symbiotic binaries (SBs) usually consist of a white dwarf accreting
material from a giant star and have orbital periods which are
typically longer than a year (sometimes many). Unlike the Cataclysmic
Variables (CVs), which have a low mass red dwarf donor, the red giant
does not generally fill its Roche Lobe and accretion onto the white
dwarf takes place through a wind. There has been speculation that SBs
might make up a significant fraction of the SN Ia progenitors
(e.g. \citealt{Kenyon1993a}). However, studies such as
\citet{Yungelson2010} and \citet{Schaefer2025} suggest that they will
not make up a significant fraction of SN Ia events.

A relatively small number of SBs show outbursts which have been
observed over a wide range of wavelengths.  These outbursts show a
considerable range in the optical amplitude, ranging from SB novae
where the amplitude is $>4$ mag, to `Z And-type' outbursts which have
amplitudes up to $\sim$2 mag. \citet{MikolajewskaKenyon1992} note
  six SB novae, whilst using a different criteria, \citet{Munari2025}
  note 11, none of which were mentioned by
  \citet{MikolajewskaKenyon1992}. Nevertheless, SB novae are fairly
  rare events but can last for a decade or more
  (e.g. \citet{Kenyon1993b}).  One of the most famous is T CrB which
showed an 8 mag amplitude outburst in 1866, with a 7 mag outburst in
1946. There has been much anticipation regarding when the next
outburst will take place (see e.g. \citealt{Merc2025a}). These
outbursts are caused by material on the white dwarf reaching a
pressure which exceeds a critical level and an explosive thermonuclear
event takes place (see \citet{MikolajewskaKenyon1992} for more
  details).

In contrast, the Z And-type outbursts have lower amplitude and are
more frequent but can also show quiescent phases lasting for more than
a decade (e.g. in Z And, \citealt{Skopal2018}). Rise times are
typically tens of days and outbursts can show complex post-peak
profiles including rebrightening events (e.g. Z And,
\citealt{Sokoloski2006} and AG Peg, \citealt{Ramsay2016}). These
  outbursts occur from instabilities in an accretion disc which
  forms from the wind of the red giant (see
  \citet{KenyonWebbink1984} for more details and \citet{Munari2019}
  for a relatively recent overview of SBs in general).

Compared to CVs, where several thousand are confirmed, with many more
candidates, there are far fewer confirmed SBs. However, SBs are
intrinsically more luminous than CVs, which allows them to be detected
in galaxies in the local group. There are 284 confirmed SBs in the
Milky Way \citep{Merc2019}, with 29 showing confirmed Z And-type
outbursts. In addition, there are 751 sources in the Milky Way that
are classed as `likely', `possible', or `suspected'. In addition,
there are 10 confirmed SBs in the LMC and 12 confirmed SBs in
the SMC. There are a further 114 sources in other galaxies such as
M31. We did not attempt to find additional sources in the
  independent study of \citet{Akras2019} who find 257 Galactic SBs and
  66 in external galaxies.

In this study we searched for outbursts in SBs in the Milky Way and
the local group using data from the GOTO all-sky survey. By expanding
the number of outburst from SBs, and characterising their profile
  and colour variation, we can provide the essential material to gain
a better understanding of the diversity of outbursts from SBs and the
physical mechanisms which power them.

\section{Observations}

We performed a systematic search for outbursts from SBs using data
obtained from the Gravitational-wave Optical Transient
Observer\footnote{\url{http://goto-observatory.org}} (GOTO), an
all-sky optical survey whose main goal is to detect the optical
counterpart of gravitational wave events
\citep{Steeghs2022,Dyer2024}. After a prototype phase, two mounts were
installed in two domes on the island of La Palma in the Canaries in
2017, each holding eight 40 cm telescopes, giving a field of view of
$\sim$44 sq deg per mount. In March 2023, two further domes were
installed at Siding Spring in Australia. The combined GOTO-North and
GOTO-South field-of-view is $\sim$176 sq deg with a pixel scale of
$\sim1.3^{''}$. The telescopes are robotically controlled and
scheduled. Observations are taken in the $L$ band filter
($\sim$400--700 nm) to a depth of $\sim$20.5--21.0 mag in stacked
images (typically 4$\times$45 sec exposures) during dark lunar
conditions.

To supplement the GOTO observations for our candidate outburst targets
we obtained the Asteroid Terrestrial-impact Last Alert System (ATLAS)
photometry using their forced photometry
tool\footnote{\url{http://fallingstar-data.com}}.  ATLAS is a
near-Earth asteroid survey which originally consisted of two
telescopes located on two separate sites in Hawaii (Mauna Loa and
Haleakala) with a single telescope being commissioned in Chile and
South Africa in 2022 \citep{Tonry2018,Henize2018}. The 0.5 m diameter
telescopes have a field of view of $\sim$29 sq deg and a pixel scale
of 1.86$^{''}$. The filters which are used are the `cyan' (420-650 nm)
and `orange' (560-820nm) bands. To determine the (c-o) colour, we used
c and o photometric points which were within 2 days of each other -
they are therefore not truly simultaneous. We also searched the
AAVSO\footnote{\url{http://aavso.org}} archive for observations which
provided additional information over the time frame of interest.

\section{Results}

We obtained the GOTO light curve of all 1213 sources in the catalogue
of SBs of \citet{Merc2019} using the GOTO forced photometry tool which
applies zero-point corrections \citep{Lyman2026} on a camera-by-camera
basis (Jarvis et al, in prep) and we used the mean $(g-r)$ colour and
$g$ mag from the PanSTARRS catalogue \citep{Chambers2016} or other
photometric measurements for calibration purposes. This query provided
photometry going back to 2023.  We then examined the resulting light
curves by-eye to identify evidence for outbursts.

Identifying recurrent novae outbursts, which are powered by nuclear
burning on the photosphere of the white dwarf, is straightforward
since they usually show amplitudes of \gtae 4 mag with a typical rise
time of $<$1 day \citep{Schaefer2010}.  There were no examples of
recurrent nova events in any of the GOTO light curves we examined.  We
therefore searched for examples of Z And-type outbursts by identifying
significant rises in brightness in SBs over a few tens of days.  Our
initial search identified ten candidate outburst events from our
targets.

The ATLAS data allow us to extend the light curves of these ten
  sources to over ten years of observations. In some cases, these
  extended observations allowed us to observe pulsations from the red
  giant star: one example is V407 Cyg which shows a previously known
  quasi-period of $\sim$700 d with a high amplitude (5--6 mag in
  ATLAS-c).  We were left with five sources which show evidence of Z
  And-type outburst behaviour.

We show the GOTO and ATLAS photometry (and in three sources with
additional AAVSO data) of the five sources which show likely outbursts
in GOTO data in Figure \ref{lightcurves-goto-atlas}. We discuss each
of these five sources in turn.

\begin{table*}
  \caption{The five SBs for which we find evidence of Z And-type
    outbursts. We show the RA, Dec co-ordinates (J2000); whether it is
    located in the Milky Way, LMC or SMC; the spectral type of the
    giant donor star; the IR type (Stellar or Dusty); whether they
    have shown previous outbursts and the orbital period if known.}
  \begin{center}
  \begin{tabular}{lrrrrrrr}
    \hline
Target & RA & DEC & Galaxy & Spectral Type & IR type & Previous Outburst? & $P_{Orb}$ \\
\hline
LMC N67 & 84.0316 & $-64.7226$ & LMC & C3.2 & S  & N & \\
OGLE SMC-LPV-4044 & +10.6995 & $-74.7000$ & SMC & G--K & S & Y & \\
HK Sco & 253.6710 & $-30.3852$ & MW & M1 III & S & Z And-type & 458$\pm$21  d (1)\\
QW Sge & 296.4565 & +18.6132 & MW &  M5 III & S & Y? & 390.5 d (2) \\
V4141 Sgr & 267.5992 & $-19.8953$ & MW & M6 III (3) & D,S &  SB Nova? & \\
\hline
  \end{tabular}
  \end{center}
  \label{table1}
  \tablebib{(1) \citet{Gromadzki2013}, (2) \citet{Munari2002}, (3)
    \citet{Merc2025b}. All other parameters from \citet{Merc2019}.}
  \end{table*}
 
\begin{figure*}
    \centering
    \includegraphics[width = 0.95\textwidth]{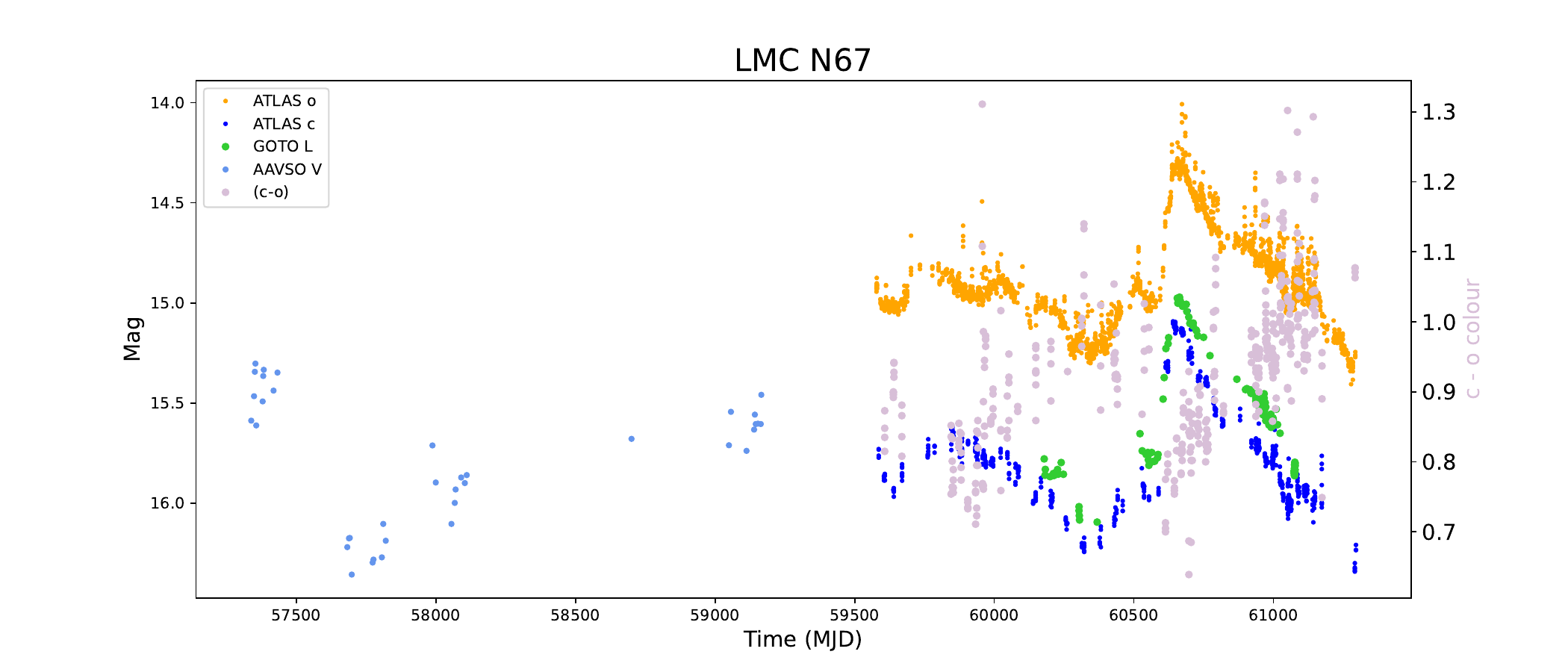}
    \includegraphics[width = 0.95\textwidth]{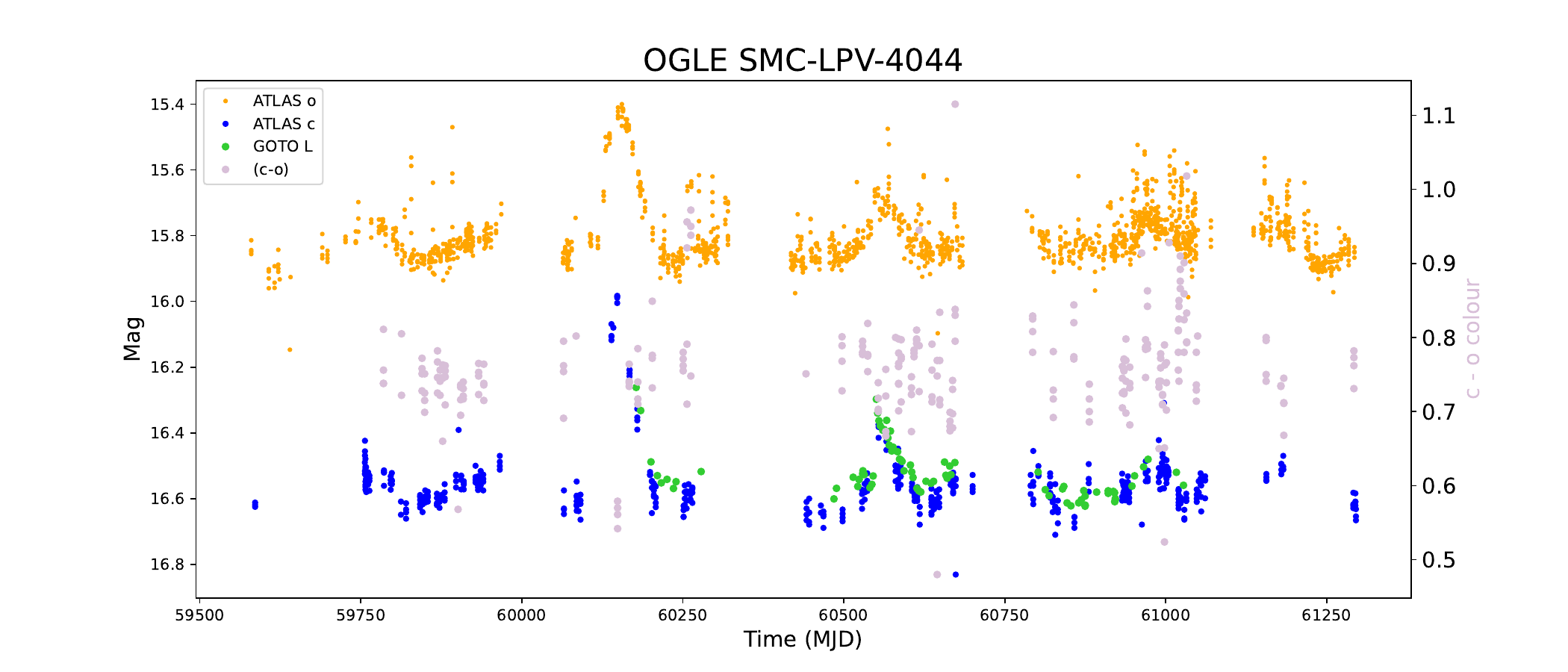}
    \includegraphics[width = 0.95\textwidth]{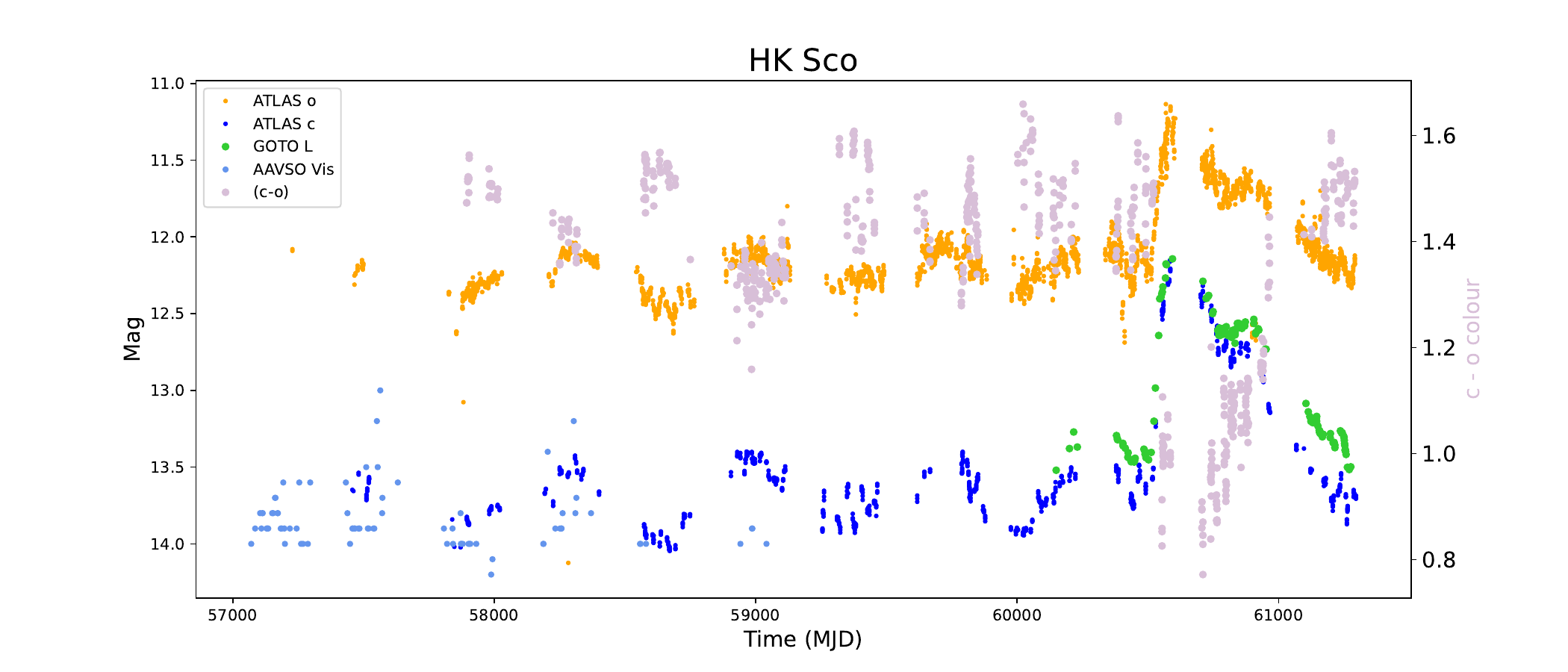}
    \caption{The light curve of Symbiotic binaries in our survey which
      showed evidence for outbursts.  GOTO L-band (green points);
      ATLAS c-band (blue); ATLAS o-band (orange); the (c-o) colour and
      data obtained from AAVSO.}
    \label{lightcurves-goto-atlas}
\end{figure*}

\setcounter{figure}{0}
 \begin{figure*}
    \centering
    \includegraphics[width = 0.95\textwidth]{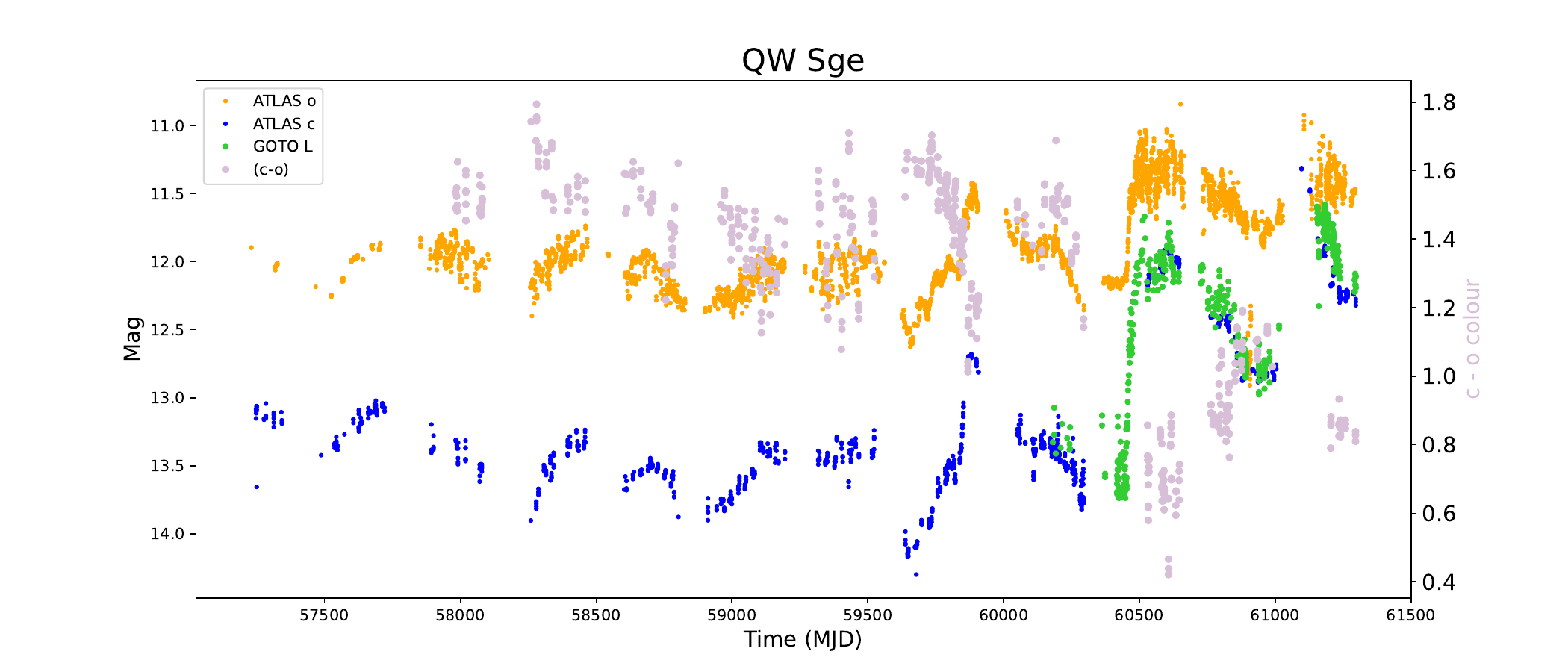}
    \includegraphics[width = 0.95\textwidth]{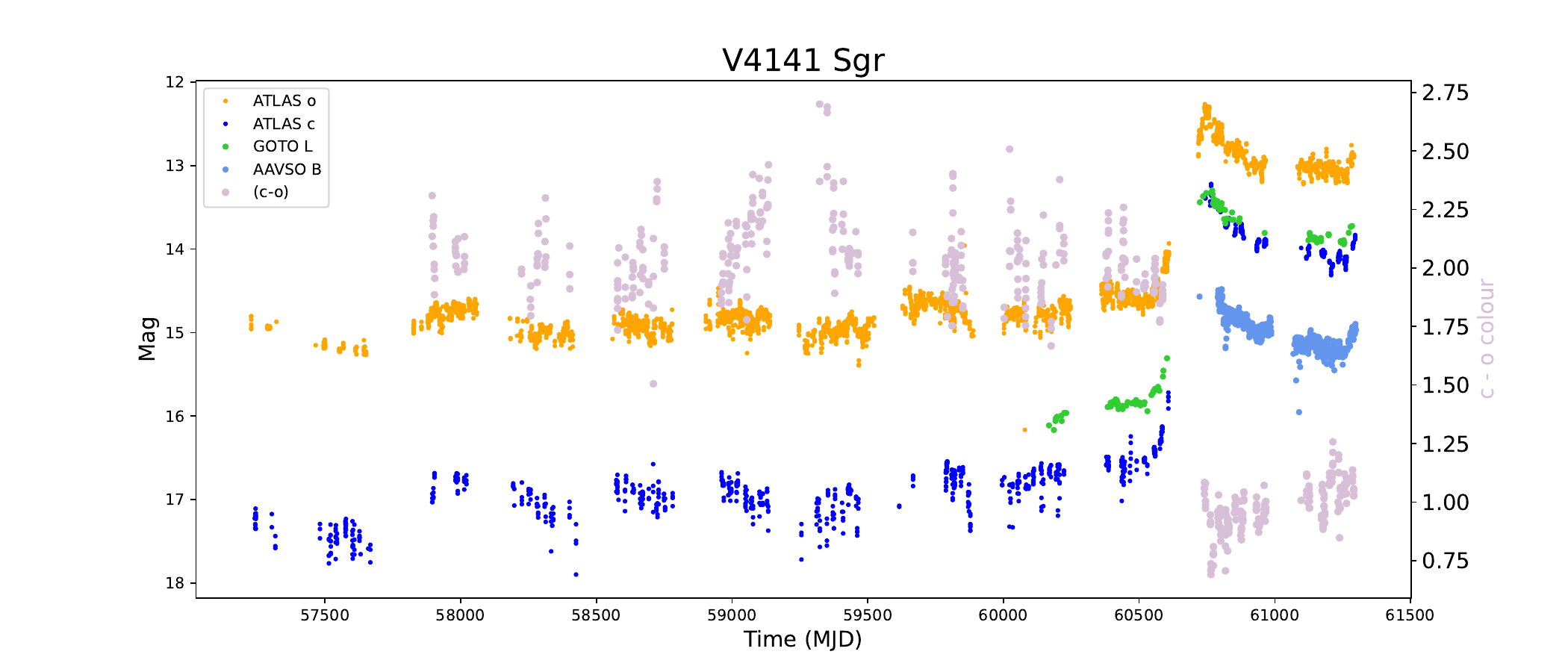}
    \caption{continued.}
\end{figure*}

\subsection{LHA 120-N 67}

LHA 120-N 67 (here after LMC N67) is included in the
\citet{Henize1956} catalogue of objects in the Magellanic Clouds which
show H$\alpha$ in emission, with \citet{Morgan1992} classifying it as
a Carbon rich SB star. Photometric observations made using OGLE showed
a mean $I$=14.5 with an amplitude of 0.33 mag \citep{Angeloni2014}.

The first ATLAS observations started in Jan 2022 with subsequent data
showing variations which have a higher amplitude in ATLAS-c. Around
2024-10-23 both GOTO and ATLAS data show a clear rise in flux reaching
a plateau $\sim$35 d later. The amplitude is 0.7 mag in ATLAS-c and
0.8 mag in both ATLAS-o and GOTO-L. By mid January 2026 there is
  a pronounced rapid decline in brightness suggesting the outburst had
  ended implying a duration of $\sim$450 d. We do not know of any
other reports of LMC N67 going into outburst.

We note there are apparent rapid brightness increases in the ATLAS-o
flux which are not seen in either the ATLAS-c or GOTO-L observations
(this discrepancy is also clearly seen in the data of OGLE
SMC-LPV-4044). We do not have any quasi-simultaneous observations of
these brightness enhancements in the GOTO-L or ATLAS-c bands. Whilst
flickering (which is thought to be due to instabilities in an
accretion disc) is quite rare in SBs, they are more commonly seen at
bluer wavelengths (see \citet{Merc2024} and references therein). It is
therefore very likely that the `flickering' seen in the ATLAS-o band
light curve is not astrophysical and could be due to the larger
  pixel scale of the ATLAS detectors and the difficulty of measuring
  the true background in crowded fields. However, the implied (c-o)
  colour variation indicates the system became noticably bluer at the
  start of the outburst, becoming redder over the course of the
  outburst.

\subsection{OGLE SMC-LPV-4044}

OGLE SMC-LPV-4044 was originally given the name SMC 2 by
\citet{Morgan1992}, and classed as an approximate G-K spectral type
and also noted there were two outbursts with amplitudes 1--1.5 mag
($B_{J}$) since 1973. The OGLE-III catalog of variable stars
\citep{Soszynski2011} reports periods of 183.1, 212.5, 276.3 d, all
with amplitudes $<0.1$ mag in the $I$ band.

The light curve shows four instances where the brightness increases by
$>$0.1 mag in the ATLAS-o band. The most prominent peaks around
MJD$\sim$60148 (late July 2023) where the amplitude is $\sim$0.5 mag
(ATLAS-o) and the rise time is shorter than the decline phase. These
brightenings have amplitudes lower than those reported by
\citet{Morgan1992}. The brightenings are separated in time by
$\sim$370, 410 and 420 d respectively. The ATLAS-o band photometry
shows more scatter than the other bands and as we noted for LMC N67,
this is unlikely to be astrophysical in origin. However, the peak
  brightness of the first candidate outburst (MJD$\sim$60150), is
  bluer, but not noticably bluer in the second, less prominent
  candidate outburst at MJD$\sim$60550).

\subsection{HK Sco}

HK Sco was identified as a star showing bright H$\alpha$ in emission by
\citet{MerrillBurwell1950}, while \citet{Allen1980} gives a spectral
type of M1. An orbital period of 458$\pm$21 d is reported by
\citet{Gromadzki2013} who also note there were some outbursts with
amplitudes up to 1.7 mag seen between 2002--2005.

The ATLAS lightcurves show evidence of a quasi-periodic modulation on
a period ($\sim$680 d) which is longer than the reported orbital
period. However, in early Aug 2024 an outburst was seen, peaking in
early Oct, (MJD$\sim$60595), with an amplitude of $\sim$1.3 mag in
ATLAS-o and GOTO-L and $\sim$1.0 mag in ATLAS-o. By early March 2026
(MJD$\sim$61105) HK Sco was approaching its pre-outburst brightness
and has continued to decline. This is the first reported outburst
since those reported by \citet{Gromadzki2013}.

Prior to the outburst we note changes in the (c-o) colour which
  are likely related to the pulsation phase of the red giant star,
  although the more rapid changes are probably due to the
  non-simultaneous nature of the observations. As seen in LMC N67, HK
  Sco is bluer at the start of the outburst, becoming redder as the
  outburst declines.

\subsection{QW Sge}

QW Sge was identified as a star showing bright H$\alpha$ in emission by
\citet{MerrillBurwell1950}, and has an orbital period of 390.5 d
\citep{Munari2002}. \citet{Allen1980} classifies the spectrum as M6
III. \citet{Munari2024} report the start of an outburst in July 2024
with previous outbursts being observed in 1963, 1983, 1997, and 2022. 

The ATLAS data show the previously reported 2022 outburst which started in 
mid Sept 2022 and peaked in mid Nov 2022 (MJD$\sim$59900) which
took place during a relatively slow rise in brightness and had an
amplitude of $\sim$0.7 mag (ATLAS-c) $\sim$0.6 mag (ATLAS-o). The
final decline of the outburst started in early Sept 2023 reaching 
near quiescence $\sim$100 d later (mid Dec 2023, MJD$\sim$60295).

The GOTO-L and ATLAS-o band light curves show the start of an outburst
around in mid-May 2024 (MJD$\sim$60450) which is $\sim$160 d after the
end of the 2022 outburst. The 2024 outburst reached a maximum
(MJD$\sim$60525) around 50 days after the start of the rise with an
amplitude of 1.4 mag (GOTO-L) and 1.0 mag (ATLAS-o). The ATLAS-o data
shows substantial variation on a night-to-night basis. There is
  evidence for a rebrightening starting towards the end of 2025, which
  continued during 2026 until May 2026 when the system started to fade
  again. As with LMC N67 and HK Sco the source is bluer during
  outburst.

\subsection{V4141 Sgr}
\label{V4141Sgr}

V4141 Sgr was originally thought to be a planetary nebula and yet was
included as a confirmed SB in the \citet{Belczynski2000}
catalogue. The mass donating star is a red giant
\citep{Merc2025b}. Those authors also report on historic photometry
including DASCH photographic plates \citep{Grindlay2012} which
indicate a slow nova outburst in the 1940's which lasted for at least
five years, and data showing an outburst starting around 1970 which
lasted for $\sim$20 years. \citet{Merc2025b} also report much more
recent observations which show an outburst with an amplitude of
$\sim$2.5--2.6 mag (ATLAS-o) starting in March 2025. The ATLAS and
GOTO data show an initial rise in brightness ($L$=15.6) in Oct 2024
reaching a maximum of $L$=13.4 early in April 2025. By mid Oct 2025,
it had reached $L$=13.8, with the most recent photometry (mid
  Sept 2026, MJD$\sim$61300) showing it reached a plateau phase.  The
  source is significantly bluer at the start of the outburst and like
  the other sources, becomes redder over the outburst.

\section{Discussion and Conclusions}

We report optical photometric observations of five SBs which show
evidence of at least one outburst. In the case of LMC N67 it is the
first recorded outburst seen in this system. We detect a number of
relatively short low amplitude outbursts in OGLE SMC-LPV-4044 and
record the first outburst in HK Sco for more than a decade. We also
report further observations of the previously noted ongoing outbursts
in QW Sge and V4141 Sgr.

With the discovery of outbursting SBs in the Magellanic Clouds,
  it allows the investigation of lower metalicity SBs to determine how
  they compare with Milky Way SBs.  OGLE-SMC-LPV-00861 was previously
  identified as being the first confirmed Z~And-type outburst in the
  SMC \citep{Miszalski2014}. The outburst of LMC~N67 now brings what
  we believe to be the first confirmed Z~And-type outburst seen in the
  LMC.  The metalicity abundance of the red giant mass donating star
  can determine the stars evolution and whether or not the star fills
  its Roche Lobe.  There are a large number of studies over many
  decades which have sought to determine the abundances of key
  elements (such as CNO and Fe). One study is that of
  \citet{Galan2016} who determined abundances of 26 Milky Way SBs
  which revealed a range of -0.8 < [Fe/H] < 0.4 dex, but with a
  general peak around -0.4 to -0.3 dex. Nine of these sources show
  outbursts according to \citet{Merc2019}. There appears to be no
  metalicity study of LMC~N67, OGLE SMC-LPV-4044 nor
  OGLE-SMC-LPV-00861 published in the literature.

In \S \ref{V4141Sgr} we report observations of V4141 Sgr which
  show a Z And-type outburst, first reported by \citet{Merc2025b}, and
  is still on-going one year after reaching its peak. This outburst is
  $\sim$35 years after the start of a slow nova \citep{Skopal2017}.
  Other examples of SBs which show slow or classical novae and also Z
  And-type outbursts include AG Peg \citep{Skopal2017}, Nova Cen 2005
  (V1047 Cen) \citep{SkopalShagatova2025}) and V426 Sge
  \citep{Skopal2020}. Accreting binaries showing both nova outbursts
  and accretion disc outbursts is not restricted to SBs. One example
  is Nova Her 1960 (V446 Her) which is a CV with an orbital period of
  4.97 hr and shows frequent dwarf nova outbursts
  \citep{Honeycutt2011}.

Compared to CVs, it is arguable that the physical processes which
  give rise to outbursts in SBs are more complex.  It is still rather
  uncertain if outbursts in the same binary could be driven by
  multiple types of physical processes. An episode of enhanced mass
  transfer could set off a disc instability which eventually triggers
  a thermonuclear runaway on the photosphere of the white dwarf
  (e.g. \citealt{Bollimpalli2018}).

By quickly identifying an outburst from a SB, multi-wavelength
observations can be made at important points in the outburst, which
can reveal details of the physics of the outburst. In four of our
  targets, the source showed a significantly bluer ATLAS (o-c) colour
  close to the point of peak outburst, becoming redder as the outburst
  declines in brightness. (The situation is less clear in SMC2 where
  the outbursts are much shorter in duration). Spectroscopic
  observations over the course of an outburst can provide more detail
  for how the physical conditions change, but multi-band photometry
  can help reveal how, for instance, the temperature of the accretion
  disc changes over time (see for instance \citet{Skopal2006}).

  They can also reveal the signature
  of the orbital period such as found in V426 Sge which showed a SB
  nova in 1968, only reaching quiescence in 1990. During its decline
  it showed a periodic modulation which was revealed as the signature
  of the orbital period \citep{Skopal2020}.  Only 45\% of confirmed
  SBs in the Milky Way have orbital period determinations which are
  one of the key parameters to define the evolution of the system.

Another set of observations which are important is detecting short
period pulsations during outbursts such as that seen in Z And during
the 1997 outburst \citep{SololoskiBildsten1999}. This signal was
interpreted as the rotation period of the white dwarf. Unexpectedly
this period was not detected during the 2000-2002 outburst, but its
absence gave clues to the physical conditions in the accretion disc
during outburst. If outbursts can be detected in near real-time, then
2--4~m class telescopes with high speed imagers would be well placed
to detect short period pulsations.

The number of ground based all-sky surveys and satellites such as TESS
\citep{Ricker2015} or Plato \citep{Rauer2025} will be able to increase
the number of outbursts detected in SBs. This will enable a better
understanding of the similarities and differences in outbursts not
only between different objects but also within systems, hence leading
to a better understanding of physical mechanisms behind the outbursts
observed in SBs.

\begin{acknowledgements}

The Gravitational-wave Optical Transient Observer (GOTO) project
acknowledges the support of the Monash-Warwick Alliance; University of
Warwick; Monash University; University of Sheffield; University of
Leicester; Armagh Observatory \& Planetarium; the National
Astronomical Research Institute of Thailand (NARIT); Instituto de
Astrofísica de Canarias (IAC); University of Portsmouth; University of
Turku; University of Birmingham; and the UK Science and Technology
Facilities Council (STFC, grant numbers ST/T007184/1, ST/T003103/1 and
ST/Z000165/1). This work has made use of data from the Asteroid
Terrestrial-impact Last Alert System (ATLAS) project which is
primarily funded to search for near earth asteroids through NASA
grants NN12AR55G, 80NSSC18K0284, and 80NSSC18K1575; by products of the
NEO search include images and catalogs from the survey area. This work
was partially funded by Kepler/K2 grant J1944/80NSSC19K0112 and HST
GO-15889, and STFC grants ST/T000198/1 and ST/S006109/1. The ATLAS
science products have been made possible through the contributions of
the University of Hawaii Institute for Astronomy, the Queen’s
University Belfast, the Space Telescope Science Institute, the South
African Astronomical Observatory, and The Millennium Institute of
Astrophysics (MAS), Chile. We gratefully acknowledge the contributions
of the AAVSO observer community, whose photometric data and metadata
resources were used in this study and made available through the
AAVSO’s scientific archives.  JDL, DON and MP acknowledge support from
a UK Research and Innovation Future Leaders Fellowship (grant
references MR/T020784/1 and UKRI1062).  TLK acknowledges support from
a Warwick Astrophysics prize post-doctoral fellowship made possible
thanks to a generous philanthropic donation.  DS acknowledges support
from The Science and Technology Facilities Council (STFC) via grants
ST/T007184/1, ST/T003103/1, ST/T000406/1, ST/X001121/1.  BW
acknowledges the UKRI’s STFC studentship grant funding, project
reference ST/X508871/1.  JC acknowledges support by the Spanish
Ministry of Science via the Plan de Generaci\'on de Conocimiento
through grant PID2022-143331NB-100.  DLC acknowledges support from the
Science and Technology Facilities Council (STFC) grant number
ST/X001121/1.  LK acknowledges support for an Early Career Fellowship
from the Leverhulme Trust through grant ECF-2024-054 and the Isaac
Newton Trust through grant 24.08(w).  AS acknowledges the Warwick
Astrophysics PhD prize scholarship made possible thanks to a generous
philanthropic donation. DMS acknowledges support through the Ram\'on y
Cajal grant RYC2023-044941, funded by MCIU/EI/10.13039/501100011033
and FSE+.  SMa acknowledges financial support from the Research
Council of Finland project 350458.  SMo is funded by Leverhulme Trust
grant RPG-2023-240.  RLCS acknowledges funding from Leverhulme Trust
grant RPG-2023-240.  This research has made use of data and/or
services provided by the International Astronomical Union's Minor
Planet Center. We thank a referee for useful comments.

\end{acknowledgements}


\vspace{4mm}


\begin{thebibliography}{99}
\bibitem[Angeloni et al. (2014)]{Angeloni2014}Angeloni, R., et al., 2014, MNRAS, 438, 35
\bibitem[Akras et al (2019)]{Akras2019}Akras, S., Guzman-Ramirez, L., Leal-Ferreira, M. L., Ramos-Larios, G.,
  2019, ApJS, 240, 21
\bibitem[Allen (1980)]{Allen1980}Allen, D. A., 1980, MNRAS, 192, 521
\bibitem[Belczy\'{n}ski et al. (2000)]{Belczynski2000}Belczy\'{n}ski, K., Mikołajewska, J., Munari, U., Ivison, R. J., Friedjung, M.,
  2000, A\&AS, 146, 407
\bibitem[Bollimpalli et al. (2018)]{Bollimpalli2018}Bollimpalli, D. A., Hameury, J.-M., Lasota, J.-P., 2018, MNRAS, 481, 5422
\bibitem[Chambers et al. (2016)]{Chambers2016}Chambers, K C., et al., 2016, arXiv-1612.05560
\bibitem[Dyer et al. (2024)]{Dyer2024}Dyer, M. J., et al., 2024, SPIE, 13094, 130941X-1
\bibitem[Ga{\l}an et al. (2016)]{Galan2016}Ga{\l}an, C., Miko{\l}ajewska, J., Hinkle, K. H., Joyce, R. R., 2016, MNRAS, 455, 1282
\bibitem[Grindlay et al. (2012)]{Grindlay2012}Grindlay, J., Tang, S., ; Los, E., Servillat, M., 2012, New Horizons in Time-Domain Astronomy, Proceedings of the International Astronomical Union, IAU Symposium, 285, 29
\bibitem[Gromadzki et al. (2013)]{Gromadzki2013}Gromadzki, M., Mikolajewska, J., Soszynski, I., 2013, AcA, 63, 405
\bibitem[Henize (1956)]{Henize1956}Henize, K., G., 1956, ApJS, 2, 315
\bibitem[Henize et al. (2018)]{Henize2018}Henize, A. N.,  et al. 2018, AJ, 156, 241
\bibitem[Honeycutt et al (2011)]{Honeycutt2011}Honeycutt, R. K., Robertson, J. W., Kafka, S., 2011, AJ, 141, 121
\bibitem[Kenyon \& Webbink (1984)]{KenyonWebbink1984}Kenyon, S. J., Webbink, R. F., 1984, ApJ, 279, 252
\bibitem[Kenyon et al. (1993a)]{Kenyon1993a}Kenyon, S. J., Livio, M., Mikolajewska, J., Tout, C. A., 1993a, ApJ, 407, L81
\bibitem[Kenyon et al. (1993b)]{Kenyon1993b}Kenyon, S. J., et al., 1993b, AJ, 106, 1573
\bibitem[Lyman et al. (2026)]{Lyman2026}Lyman, J. D., et al., 2026, submitted to RASTI, arXiv:2603.02330 
\bibitem[Merc et al. (2019)]{Merc2019}Merc, J., G'{a}lis, R., Wolf, M., 2019, RNAAS, 3, 28
\bibitem[Merc et al. (2024)]{Merc2024}Merc, J., et al., 2024, A\&A, 683, A84
\bibitem[Merc et al. (2025a)]{Merc2025a}Merc, J., et al., 2025, MNRAS, 541, L1
\bibitem[Merc et al. (2025b)]{Merc2025b}Merc, J., et al., 2025, A\&A, 698, L4
\bibitem[Merrill \& Burwell (1950)]{MerrillBurwell1950}Merrill, P. W., Burwell, C. G., 1950, ApJ, 111, 72
\bibitem[Mikolajewska \& Kenyon (1992)]{MikolajewskaKenyon1992}Mikolajewska, J. \& Kenyon, S. J., 1992, MNRAS, 256, 177
\bibitem[Miszalski et al (2014)]{Miszalski2014}Miszalski, B., Mikolajewska, J., Udalski, A., 2014, A\&A, 444, 11
\bibitem[Morgan (1992)]{Morgan1992}Morgan, D. H., 1992, MNRAS, 258, 639
\bibitem[Munari \& Jurdana-{\v{S}}epi{\'c} (2002)]{Munari2002}Munari, U., Jurdana-{\v{S}}epi{\'c}, R., 2002, A\&A, 386, 237
\bibitem[Munari (2019)]{Munari2019}Munari, U., 2019, Invited Review, published in "The Impact of Binary Stars on Stellar Evolution",
G. Beccari and M.J. Boffin eds., Cambridge Univ. Press, Cambridge Astrophysical Series vol. 54, 77,
  Proceedings of a Conference held on 3-7 July 2017, at ESO Headquarters, Garching, Germany
\bibitem[Munari (2024)]{Munari2024}Munari, U., Righetti, G. L., Farina, A., 2024, ATel, 16739 
\bibitem[Munari (2025)]{Munari2025}Munari, U., 2025, Contributions of the Astronomical Observatory Skalnat\'{e} Pleso, 2025, 55, 47 
\bibitem[Ramsay et al. (2016)]{Ramsay2016}Ramsay, G.,  et al., 2016, MNRAS 461, 3599
\bibitem[Rauer et al. (2025)]{Rauer2025}Rauer, H., et al., 2025, Experimental Astronomy, 59, 26
\bibitem[Ricker et al. (2015)]{Ricker2015}Ricker G. et al., 2015, JATIS, 1a4003
\bibitem[Schaefer (2010)]{Schaefer2010}Schaefer, B. E., 2010, ApJS, 187, 275
\bibitem[Schaefer (2025)]{Schaefer2025}Schaefer, B. E., 2025, ApJ, 994, 187
\bibitem[Skopal et al. (2006)]{Skopal2006}Skopal, A., et al., 2006, A\&A, 453, 279
\bibitem[Skopal et al. (2017)]{Skopal2017}Skopal, A., et al., 2017, A\&A, 604, A48
\bibitem[Skopal et al. (2018)]{Skopal2018}Skopal, A., Tarasova, T. N., Wolf, M., Dubovsk'{y}, P. A., Kudzej, I., 2018, ApJ, 858, 120
\bibitem[Skopal et al. (2020)]{Skopal2020}Skopal, A., et al., 2020, A\&A, 636, A77
\bibitem[Skopal \& Shagatova (2025)]{SkopalShagatova2025}Skopal, A., Shagatova, N., 2025, ApJ, 983, 143
\bibitem[Sokoloski \& Bildsten (1999)]{SololoskiBildsten1999}Sokoloski, J. L., Bildsten, L., 1999, ApJ, 517, 919
\bibitem[Sokoloski et al. (2006)]{Sokoloski2006}Sokoloski, J. L., et al., 2006, ApJ, 636, 1002
\bibitem[Soszynski et al. (2011)]{Soszynski2011}Soszynski, I., et al. 2011, AcA, 61, 217
\bibitem[Steeghs et al. (2022)]{Steeghs2022}Steeghs, D., et al., 2022, MNRAS, 511, 2405
\bibitem[Tonry et al. (2018)]{Tonry2018}Tonry, J. L., et al. 2018, PASP, 988, 064505
\bibitem[Yungelson (2010)]{Yungelson2010}Yungelson, L. R., 2010, AstL, 36, 780
  
\end{thebibliography}
\end{document}